# Superconductivity in the medium-entropy alloys: $Nb_2TiW$ and $Nb_2TiMo$

Kuan Li[a,#], Cui-Qun Chen[b,#], Lingyong Zeng[a], Longfu Li[a], Rui Chen[a], Peifeng Yu[a], Kangwang Wang[a], Zaichen Xiang[a], Dao-Xin Yao[b,c,d,*], Huixia Luo[a, c,d,*]

[a] School of Materials Science and Engineering, Sun Yat-sen University, No. 135, Xingang Xi Road, Guangzhou, 510275, P. R. China

[b] School of Physics, Sun Yat-sen University, No. 135, Xingang Xi Road, Guangzhou, 510275, P. R. China

[c] State Key Laboratory of Optoelectronic Materials and Technologies, Sun Yat-sen University, No. 135, Xingang Xi Road, Guangzhou, 510275, P. R. China

[d] Guangdong Provincial Key Laboratory of Magnetoelectric Physics and Devices, Sun Yat-sen University, No. 135, Xingang Xi Road, Guangzhou, 510275, P. R. China

[#] K. Li and C. Chen contributed equally to this work.

*Corresponding author/authors complete details (Telephone; E-mail:) (+86)-2039386124, E-mail address: yaodaox@mail.sysu.edu.cn; luohx7@mail.sysu.edu.cn

**Abstract**

This study describes the synthesis and characterization of $Nb_2TiW$ and $Nb_2TiMo$ medium-entropy alloys (MEAs). The $Nb_2TiW$ and $Nb_2TiMo$ MEAs can be successfully synthesized by an arc-melting method. Their structures and superconducting properties are investigated by detailed characterization of X-ray diffraction (XRD), resistivity, magnetization, and specific heat measurements. XRD results confirm that the obtained $Nb_2TiW$ and $Nb_2TiMo$ compounds have the same body-centered cubic (BCC) structures and crystalize in the $Im\bar{3}m$ space group (number 229). Experimental results show that the superconducting transition temperatures $T_c$s of $Nb_2TiW$ and $Nb_2TiMo$ are around 4.86 K and 3.22 K, respectively. The upper and lower critical fields of $Nb_2TiW$ are 3.52(2) T and 53.36(2) Oe, respectively, and those of $Nb_2TiMo$ are 2.11(2) T and 68.23(3) Oe, respectively. First-principles calculations reveal that the $d$ electrons of Nb, Ti, and Mo or W are the dominant contribution of the density of states near the Fermi level. Specific heat measurement results indicate that $Nb_2TiW$ and $Nb_2TiMo$ display BCS full-gap $s$-wave superconductivity.

Medium-entropy alloys (MEAs) and high-entropy alloys (HEAs) are a new class of metallic materials, the study of which has now developed into a cutting-edge direction involving the cross-fertilization of multiple disciplines such as materials, physics, chemistry, mechanics, and computational science.[1-9] Their unique atomic arrangement, crystal structure, and interaction mechanism have received significant attention in materials science, catalyst design, energy storage, and other fields.[10-12]

One of the most attractive properties of MEAs and HEAs in condensed matter physics is superconductivity. These solid solutions of MEAs and HEAs consist of multiple components and are characterized by a significant mixing entropy ($\Delta S_{mix}$). One definition is based on the magnitude of the mixing entropy $\Delta S_{mix}$, $\Delta S_{mix}$ is calculated by the following equation: $\Delta S_{mix} = -R \sum_{i=1}^{n} c_i \ln c_i$, where $n$ is the number of components, $c_i$ is the atomic fraction, and $R$ is the gas constant. According to this equation, the $\Delta S_{mix}$ of MEA is between 0.69 $R$~1.60 $R$. As the number of constituent elements in the solid solution increases, the $\Delta S_{mix}$ enhances accordingly. Depending on the magnitude of the desired $\Delta S_{mix}$, low-entropy alloys are classified as those with $R$-values less than 0.69, whereas the $R$-values of HEAs are 1.60 or higher.[13] MEAs and HEAs are composed of three or more elements combined in the range of 5 - 35 % atomic.[14, 15] These MEAs and HEAs produce more compositional disorder (high $\Delta S_{mix}$) than conventional alloys due to the difference in atomic size.[16] The $Ta_{34}Nb_{33}Hf_8Zr_{14}Ti_{11}$ HEA with a superconducting transition temperature of $T_c \approx 7.3$ K was the first high-entropy superconductor to be reported.[17] Since then, more and more MEA and HEA superconducting materials have been discovered.[18-21] To date, MEAs and HEAs superconductors possess many types of crystal structures, including body-centered cubic (BCC), face-centered cubic (FCC), hexagonal close-packed (HCP), CsCl-type, $\alpha$-Mn-type[22-24] and so on. Recently, some MEAs superconducting materials have been developed and showed unique superconducting properties. The BCC-type TiHfNbTa MEA, for example, presents extremely coupled $s$-wave superconductivity [19]. In addition, TiZrHfNb MEA displays record-high $T_c$ and dome-shaped superconductivity under applied pressure conditions [25]. Another HCP-type MoReRu MEA superconductor with a valence electron count (VEC = 7) has a $T_c = 9.1$ K [26].

Previous reports demonstrate that the VEC of MEA superconductors plays a crucial role in determining their $T_c$s, as well as the stability of the material [23, 27, 28]. On the other hand, the element Nb content also affects the superconducting temperature. Remarkably, the element Nb is the single element with the highest $T_c$ at atmospheric pressure, with 9.23 K [19]. Therefore, the element Nb is commonly used in the design of superconducting materials for MEAs and HEAs. In addition, NbTi alloy is a well-known commercial superconducting material with excellent properties such as accessible $T_c$, high critical current density, high critical magnetic field, and easy processing [29-33]. Moreover, the $T_c$ of NbTi alloys increases from ~ 9.6 K to ~ 19.1 K at high pressure. NbTi alloys are the most robust of the known stressed superconductors [34]. More recently, Nb$_2$TiW and Nb$_2$TiMo with the $Fm\bar{3}m$ structure was predicted to be superconducting and highly ductile, making them potential candidates for the fabrication of wires and strips for superconducting magnets [35]. By machine learning prediction, the Nb$_2$TiW compound shows the advantage of a large critical temperature, $T_c^{\text{Eliashberg}}$ = 11 K ($T_c^{\text{McMillan}}$ = 9 K) and energy above the convex hull ($E_{\text{hull}}$) = 25 meV/atom. Another compound, Nb$_2$TiMo is also at the lower of ductility hyperbole $(G/B = 0.2$ and $(C_{12} - C_{44}) / E = 1.0)$, with the advantage of being more energy stable, only 25 meV/atom from the shell.

Based on the considerations mentioned above, we report the synthesis, crystal structure, and physical properties of Nb$_2$TiW and Nb$_2$TiMo compounds with VEC = 5 e/a. The $\Delta S_{mix}$ of the new MEA superconductors Nb$_2$TiW and Nb$_2$TiMo are $\Delta S_{mix}$ = 1.040 R, which lie between 0.69 and 1.60, and thus both are categorized as MEA. Our XRD results confirm that obtained Nb$_2$TiW and Nb$_2$TiMo MEA are adopted in the BCC structure, while it is different from the $Fm\bar{3}m$ structure used in prediction [35]. The $T_c$s of Nb$_2$TiW and Nb$_2$TiMo are 4.86 K and 3.22 K, respectively, as determined from electrical resistivity, magnetization, and specific heat measurements. Our specific heat measurements indicate that these two Nb$_2$TiW and Nb$_2$TiMo MEA are BCS full-gap $s$-wave superconductors.

The MEA Nb$_2$TiW and Nb$_2$TiMo compounds were synthesized by an extraordinary arc-melting method. The raw materials for the Nb$_2$TiW MEA were niobium from Macklin (99.99%, 325 mesh), titanium from Macklin (99.99%, 300 mesh), and tungsten from Macklin (99.5%, 325 mesh). Similarly, for the Nb$_2$TiMo MEA, we chose niobium from Macklin (99.99%, 325 mesh), titanium from Macklin (99.99%, 300 mesh), and molybdenum from Alfa Aesar (99.9%, 250 mesh). To begin the synthesis process, we carefully weighed a mixture of 250 mg of elemental powders in a precise 2:1:1 molar ratio. Then, a sample of all the powders was placed in a mortar and thoroughly ground for 1 hour to make a homogeneous mixture. We carefully pressed a pellet from this mixture. The pellet was transferred to an electric arc furnace and melted under a controlled argon environment at 0.5 atm. To ensure homogeneity throughout the material, we perform a series of turning and remelting procedures on the ingot to blend it to perfection. In the end, we obtained pure $Nb_2TiW$ and $Nb_2TiMo$ MEA materials.

The phase purity of the obtained compounds was characterized by powder X-ray diffraction (XRD). XRD data are collected with great precision, with an angular range of 10 ° to 100 °, in steps as small as 0.01 °. For the analysis, we use a Japanese Rigaku MiniFlex instrument equipped with Cu K$\alpha$ radiation by continuously scanning at a constant rate of 1 °/min. To refine and elucidate the XRD patterns, we used the Rietveld method, supplemented by the comprehensive FULLPROF software suite [36]. In addition, we used backscattered scanning electron microscopy (BSEM) and scanning electron microscope (SEM) to fully characterize the elemental proportions and microstructure of the $Nb_2TiW$ and $Nb_2TiMo$ materials (EVO, Zeiss). In addition, the EVO microscope has an integrated energy dispersive X-ray spectroscopy (EDXS) capability, which allowed us to analyze the elemental composition and distribution of the samples, thereby deepening our understanding of the sample microstructures. To gain insight into the superconducting behavior of the material, we carried out the resistivity using the physical property measurement system (PPMS, DynaCool-14T, Quantum Design). Our obtained samples were tested from 1.8 K to 300 K using a four-point probe technique. In addition to resistivity, we also reveal the magnetization properties through zero-field-cooling (ZFC) and field-cooling (FC) magnetization experiments. Finally, to understand the bulk nature of the superconducting properties of the samples, we also measured the specific heat of the material over the temperature range of 1.8 K to 15 K.

The density functional theory (DFT) calculations were performed using the Vienna ab initio simulation package (VASP) [37, 38]. The generalized gradient approximation (GGA) with the Perdew-Burke-Ernzerhof (PBE) exchange-correlation function was used in the DFT calculations [39]. The plane wave cutoff energy was set to be 600 eV, and a k-mesh grid of 9 ×9 ×9 centered at the Γ-point was sampled in the Brillouin zone. The total energy criterion at the self-consistent step and force convergences criteria were $10^{-6}$ eV and 0.001 eV/Å, respectively. The experimentally measured lattice constants of the MEAs were adopted in the DFT calculations. Spin-orbit coupling (SOC) was included in band structure calculations. To describe the chemical disorder in the MEAs, the 2 ×2 ×2 supercells special quasi-random structures (SQSs) with 16 atoms were utilized, which are modeled by Monte Carlo Special Quasi-random Structure as implemented in the Alloy Theoretic Automated Toolkit [40, 41].

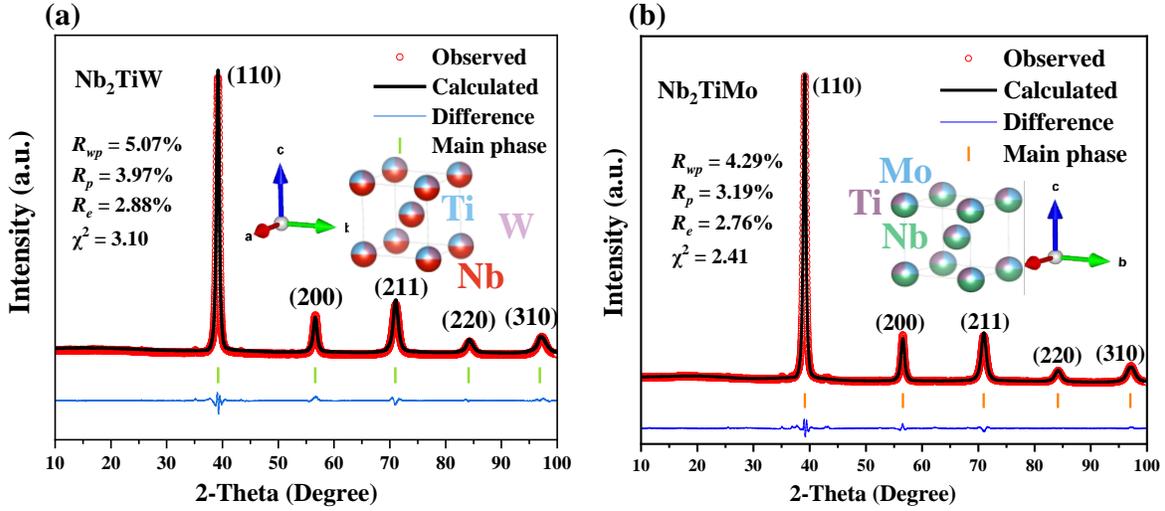

Fig. 1. (a) XRD refinement of Nb$_2$TiW. The inset shows the crystal structure of Nb$_2$TiW MEA alloy with $Im\bar{3}m$ space (b) XRD refinement of Nb$_2$TiMo. The inset shows the crystal structure of Nb$_2$TiMo MEA alloy with $Im\bar{3}m$ space.

The X-ray diffraction (XRD) spectra of the MEA Nb$_2$TiW and Nb$_2$TiMo superconducting samples and the results of their correlation analyses are presented in Fig. 1(a) and Fig. 1(b), respectively. Based on the XRD spectra, the products can be identified as pure Nb$_2$TiW and Nb$_2$TiMo superconducting materials. For ease of labeling, all observed XRD peaks were labeled with their respective Miller indices. In addition, the mild expansion of the peak signals is due to the highly disordered structure. It was relatively easy to index the XRD data in the BCC structure, where the space group $Im\bar{3}m$ and the identification number No. 229 were used. It is different from the previous prediction using pace group $Fm\bar{3}m$ (No. 225). Fitting the XRD profiles by Rietveld refinement, the values of the refinement parameters $R_{wp}$, $R_p$, $R_e$, and $\chi^2$ for the MEA Nb$_2$TiW samples were 5.07 %, 3.97 %, 2.88 %, and 3.10, respectively. The values of refinement parameters $R_{wp}$, $R_p$, $R_e$, and $\chi^2$ for the MEA Nb$_2$TiMo samples were 4.29 %, 3.19 %, 2.76 % and 2.41, respectively. These fitting data indicate that the fit is good. The final lattice parameters of the MEA Nb$_2$TiW and Nb$_2$TiMo crystals were obtained as $a = b = c = 3.248(11)$ Å and $a = b = c = 3.247(17)$ Å. The crystal structures of the Nb$_2$TiW and Nb$_2$TiMo superconducting samples are shown in the insets of Fig. 1(a) and Fig. 1(b), respectively. It can be seen that both compounds have BCC crystal structures, and the three elemental atoms Nb, Ti, and W in Nb$_2$TiW randomly occupy the lattice sites. In addition, the homogeneity and chemical composition of the Nb$_2$TiW and Nb$_2$TiMo samples were characterized using SEM-EDXS. Fig. S1(a) and Fig. S1(b) present the ratios of the constituent elements, which, according to the accurate chemical formulae, are $Nb_{2.3}Ti_{0.8}W_{0.9}$ and $Nb_{1.9}Ti_{1.2}Mo_{0.9}$. The actual elemental ratios are very close to the design values, suggesting that the prepared samples are as expected. The minor deviations may be caused by the limited precision of EDXS and the unevenness of the sample surface. From the SEM and BSEM images in Fig. 2, as well as the elemental mapping using EDX, we can observe that the studied $Nb_2TiW$ and $Nb_2TiMo$ compounds are homogeneous under the microscope. These results reveal the crystal structure, chemical composition, and homogeneity characteristics of the $Nb_2TiW$ MEA and $Nb_2TiMo$ MEA superconducting samples, which provide the basis and foundation for confirming the purity and the actual elemental content of the MEA $Nb_2TiW$ and $Nb_2TiMo$ superconducting samples.

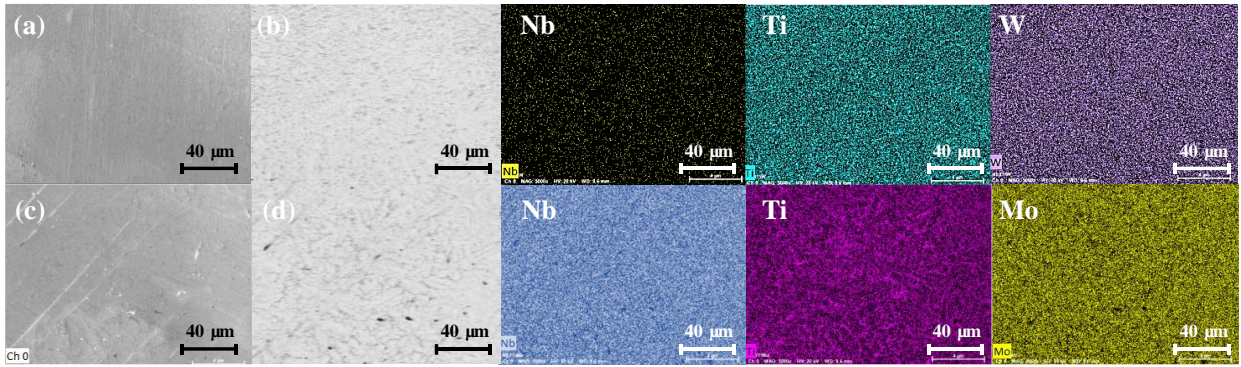

Fig. 2. (a) SEM image mapping of the $Nb_2TiW$ MEA, (b) BSEM spectrum mapping of the $Nb_2TiW$ MEA, and EDX elemental mappings of the $Nb_2TiW$ MEA. (c) SEM image mapping of the $Nb_2TiMo$ MEA. (d) BSEM spectrum mapping of the $Nb_2TiMo$ MEA and EDX elemental mappings of the $Nb_2TiMo$ MEA.

Resistivity testing is a commonly used experimental method to evaluate the performance characteristics of superconducting materials. In this study, we show the variation of normalized resistivity ($\rho/\rho_{300\,K}$) in the temperature range of 1.8 ~ 300 K for $Nb_2TiW$ and $Nb_2TiMo$ MEA samples in Fig. 3(a) and Fig. 3(c), respectively. For the $Nb_2TiW$ MEA sample, the resistivity exhibits metallic behavior as the temperature decreases from 300 K to about 6 K. The resistivity of the samples is also shown in the normalized resistivity ($\rho/\rho_{300\,K}$) for the $Nb_2TiW$ MEA sample. For the $Nb_2TiMo$ MEA sample, the resistivity similarly exhibits metallic behavior as the temperature decreases from 300 K to about 4 K. The resistivity of the $Nb_2TiMo$ MEA sample shows a metallic behavior at lower temperatures. However, at much lower temperatures, the resistivity drops dramatically and tends to zero, clearly showing the presence of superconducting properties. The residual resistivity ratio (*RRR*) is a key parameter in our research as it is an indicator of the electronic properties and quality of the material under study [42-44]. In general, higher *RRR* values indicate lower levels of impurities and defects, which are essential to ensure structural homogeneity and excellent electrical conductivity in metallic systems [43]. Here, our calculated *RRR* value is the ratio of the resistivity of the samples at 300 K and 5 K, which is used to evaluate the purity and conductivity of the samples. We calculated the RRR = $R_{300K}/R_{5K}$. For the $Nb_2TiW$ superconductor, we measured its resistivity values to be 32.91 μΩ-cm at 300 K and 21.78 μΩ-cm at 5 K. The resulting *RRR* value for $Nb_2TiW$ is therefore $R_{300K}/R_{5K}$ = 1.51. Similarly, for $Nb_2TiMo$ the resistivity value is 144.92 μΩ-cm at 300 K and 99.38 μΩ-cm at 5 K. Therefore, the *RRR* value for $Nb_2TiMo$ is $R_{300K}/R_{5K}$ = 1.46. Both of them are relatively low values, comparable to that observed for nonstoichiometric or highly disordered intermetallic compounds [23, 27, 28].

To describe the superconducting behavior of the two materials more specifically, Fig.3(b) and Fig. 3(d) demonstrate the resistivity trend over the temperature range of 1.8 ~ 10 K and 1.8 ~ 5 K, respectively. The resistivity of the two materials is shown in Fig. 3(a) and Fig. 3(c). We determined the $T_c$ using a 50 % reduction in resistivity relative to the value at the normal state. By analyzing the data, we determined the $T_c$ of the $Nb_2TiW$ and $Nb_2TiMo$ MEA superconductors to be about 4.86 K and 3.22 K, respectively. It is noteworthy that the resistivity of the $Nb_2TiW$ MEA superconductor reaches zero at about 4.45 K. Also, at about 3.13 K, the resistivity of the $Nb_2TiMo$ MEA superconductor reaches zero. These indicate the presence of its superconducting nature. It is worth noting that the range over which the resistivity undergoes drastic changes is very narrow. These results are crucial for our in-depth understanding and evaluation of the superconducting properties and performance characteristics of the $Nb_2TiW$ and $Nb_2TiMo$ MEA superconductors and provide a valuable reference for future applications and research of superconducting materials.

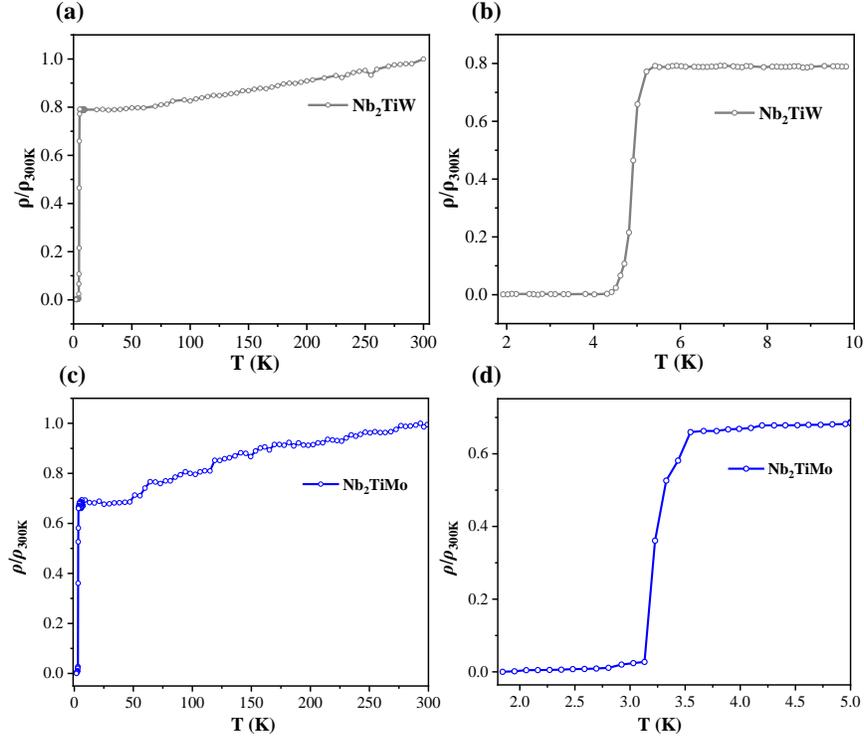

Fig. 3. (a) Temperature dependence of normalized $\rho/\rho_{300K}$ of the Nb$_2$TiW sample. (b) Low-temperature behavior of Nb$_2$TiW MEA near $T_c$. (c) Temperature dependence of normalized $\rho/\rho_{300K}$ of the Nb$_2$TiMo sample. (d) Low-temperature behavior of Nb$_2$TiMo MEA near $T_c$.

Fig. 4(a) and Fig. 4(b) show the magnetization versus temperature of the Nb$_2$TiW MEA and Nb$_2$TiMo MEA superconductor under an applied magnetic field of 30 Oe in the temperature range of 1.8 - 8 K and 1.8 - 6 K, respectively. The measurements were performed under ZFC and FC. The diameters and heights of the two cylindrical samples Nb$_2$TiW and Nb$_2$TiMo are 3 mm×4 mm and 2.5 mm×3 mm, respectively. The demagnetization factors are corrected according to the shape of the cylindrical samples by using the corresponding demagnetization factor formula, $N^{-1} = 1 + 1.6\frac{c}{a}$. The demagnetization factors of Nb$_2$TiW and Nb$_2$TiMo are about 0.3[45]. The trend of rapid decrease of magnetization with temperature in Nb$_2$TiW MEA and Nb$_2$TiMo MEA superconductors is confirmed by the strong diamagnetic signals observed at $T_c$ = 4.66 K and $T_c$ = 3.21 K, which confirm the presence of superconductivity in Nb$_2$TiW and Nb$_2$TiMo, respectively. To obtain a more accurate magnetization profile, the isothermal magnetization M(H) data were measured at 1.8 K. Fig. 4(c) and Fig. 4(d) shows the magnetic isotherm measured in the temperature range of 1.8 - 3.6 K for Nb$_2$TiW and 1.8 - 2.6 K for Nb$_2$TiMo. Under the assumption of a perfect magnetic field response, the equation for the linear fit in the low magnetic field region is $M_{fit} = m + nH$, where $m$ is the intercept and $n$ is the slope of the straight line [46]. Fig. S2 illustrates the $M-M_{fit}$ curves at different temperatures. Usually, the value of $\mu_0 H_{c1}(0)$ is extracted when the difference between $M$ and $M_{fit}$ exceeds 1% $M_{max}$. The obtained points shown in the main panels of Fig.4(c) and Fig. 4(d) fit well into the empirical formula: $\mu_0 H^*_{c1}(T) = \mu_0 H^*_{c1}(0)(1-(T/T_c)^2)$, where $T_c$ is the superconducting transition temperature and $\mu_0 H_{c1}(0)$ is the lower critical field at 0 K. Thus, the lower critical magnetic field $\mu_0 H_{c1}(0) = 53.36(2)$ Oe for Nb$_2$TiW and $\mu_0 H_{c1}(0) = 68.23(3)$ Oe for Nb$_2$TiMo can be calculated exactly.

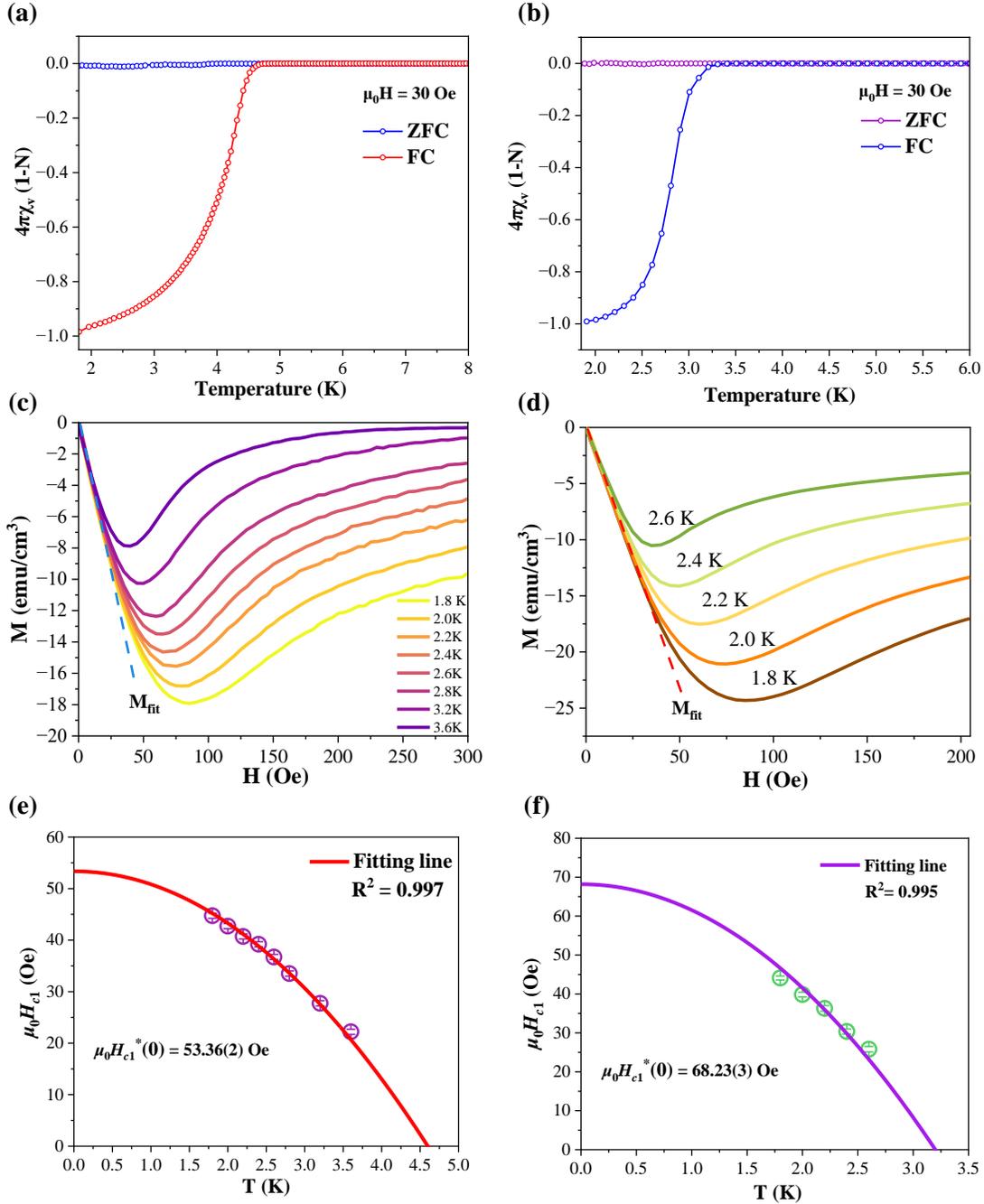

**Fig. 4.** (a) and (b) Magnetic susceptibilities for Nb$_2$TiW and Nb$_2$TiMo at the superconducting transitions, applied DC fields are 30 Oe. (c) and (d) The isothermal magnetization curve of the Nb$_2$TiW within the range of 1.8 to 3.6 K and Nb$_2$TiMo superconductor within the range of 1.8 to 2.6 K. (e) and (f) The temperature dependence of the effective lower critical field of the Nb$_2$TiW and Nb$_2$TiMo superconductor, respectively.

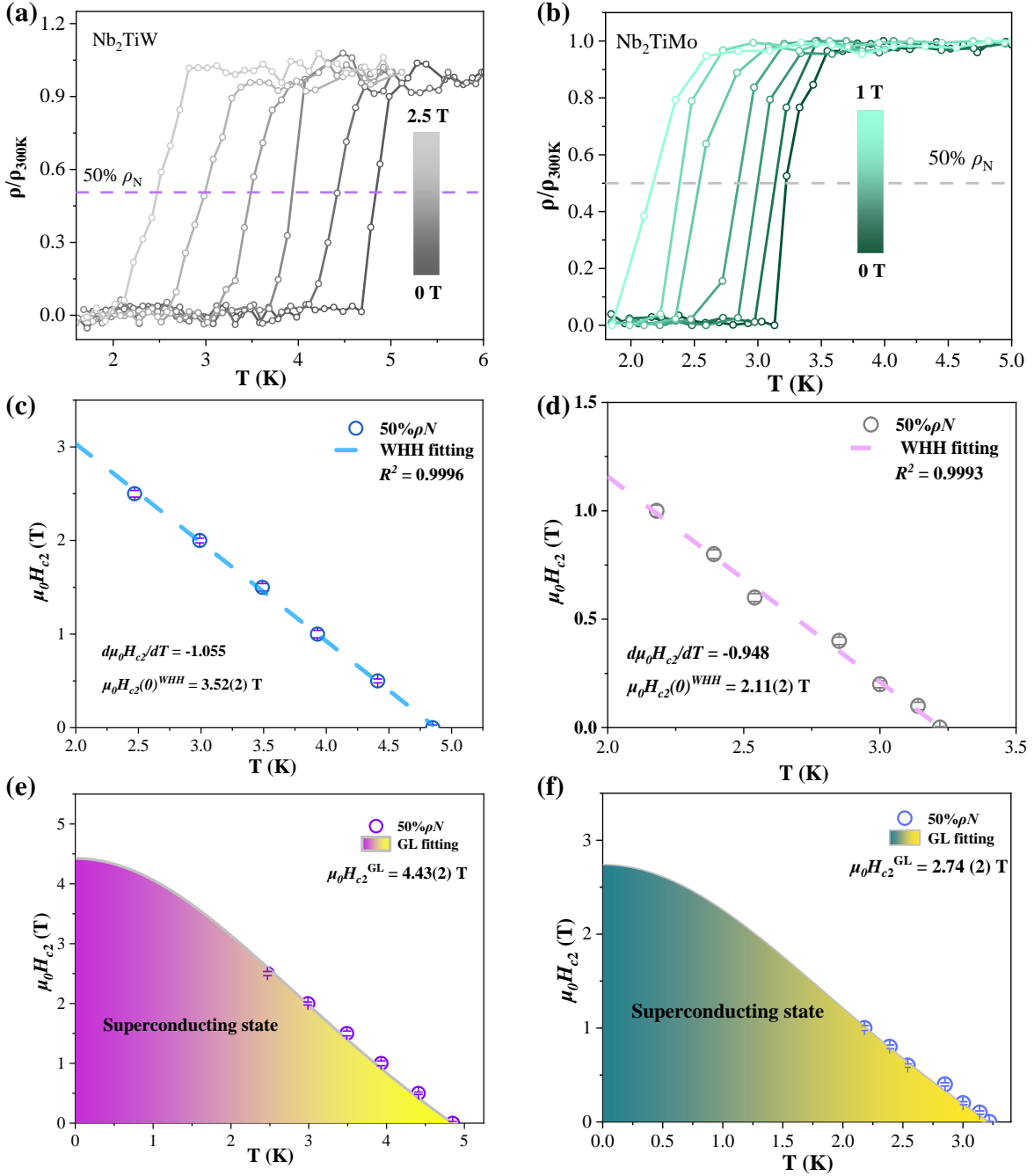

Fig. 5. (a) The resistivity of Nb$_2$TiW MEA under superconducting temperatures with the applied magnetic fields from 0 to 2.5 T. (b) The resistivity of Nb$_2$TiMo MEA under superconducting temperatures with the applied magnetic fields from 0 to 1.0 T. (c) The temperature dependence of the $\mu_0H_{c2}$ and the WHH model fitting of Nb$_2$TiW MEA. (d) The temperature dependence of the $\mu_0H_{c2}$ and the WHH model fitting of Nb$_2$TiMo MEA. (e) The $\mu_0H_{c2}$-T phase diagram of the Nb$_2$TiW MEA superconductor. (f) The $\mu_0H_{c2}$-T phase diagram of Nb$_2$TiMo MEA superconductor.

In addition, to gain a deeper understanding of the superconducting properties of Nb$_2$TiW and Nb$_2$TiMo MEA, the resistivity data of Nb$_2$TiW and Nb$_2$TiMo MEA superconductors were investigated using the R-T measurement system at various fixed magnetic fields. Fig. 5(a) demonstrates the low-temperature resistivity of Nb$_2$TiW MEA at different magnetic fields in the magnetic field range 0 - 2.5 T. Meanwhile, Fig. 5(b) depicts the low-temperature resistivity of Nb$_2$TiMo MEA under the magnetic field magnetic field range of 0 - 1.0 T. At low temperatures, SC exhibits a sharp decrease in $\rho$(T) down to the value of $\rho = 0$. When $T_c$ is defined as a 50% decrease in resistivity related to the normal state value, the transition temperatures of the $\rho$(T) data for Nb$_2$TiW and Nb$_2$TiMo MEA superconductors are categorized as $T_c$ = 4.86 K and $T_c$ = 3.22 K. When the magnetic field is increased, the $T_c$ shifts to lower temperatures and the superconducting transition becomes wider. We obtained the temperature variation of $\mu_0H_{c2}(0)$ for Nb$_2$TiW and Nb$_2$TiMo MEA samples. The slope ($d\mu_0H_{c2}/dT$) of the Nb$_2$TiW MEA compound is -1.055 T/K, as shown in Fig. 5(c). The slope ($d\mu_0H_{c2}/dT$) of the Nb$_2$TiMo MEA compound is -0.948 T/K, as shown in Fig. 5(d). The data were fitted with the WHH formula: $\mu_0H_{c2}(0) = -0.693T_c(\frac{d\mu_0H_{c2}}{dT})|_{T=T_c}$. Using $T_c$ = 4.86 K (50% $\rho_N$), the dirty limit $\mu_0H_{c2}(0)$ WHH value for the Nb$_2$TiW MEA superconductor was calculated to be 3.52(2) T. Using the same fit with the WHH formula, using $T_c$ = 3.22 K (50% $\rho_N$), the dirty limit $\mu_0H_{c2}(0)$ WHH value for the Nb$_2$TiMo MEA superconductor was calculated to be 2.11(2) T. In order to explore the properties of Nb$_2$TiW and Nb$_2$TiMo MEA more precisely, we defined the resistivity value as low as 50% of $\rho_N$ as $\mu_0H_{c2}$ and calculated $\mu_0H_{c2}(0)$ using the Ginzburg-Landau (GL) equation: $\mu_0H_{c2}(T) = \mu_0H_{c2}(0) \times \frac{(1-(T/T_c)^2)}{(1+(T/T_c)^2)}$. Fitting the data yields $\mu_0H_{c2}(0)^{GL}$ = 4.43(2) T for Nb$_2$TiW and $\mu_0H_{c2}(0)^{GL}$ = 2.74(2) T for Nb$_2$TiMo MEA. Thus, the phase diagrams of $\mu_0H_{c2}(0)$ relative to T for Nb$_2$TiW and Nb$_2$TiMo MEA are shown in Fig. 5(e) and Fig. 5(f). The GL model satisfactorily fits the experimental data over the entire temperature range. It is worth noting that the upper critical field calculated by the WHH model and the GL model must be smaller than the Pauli limit field. The $\mu_0H^{Pauli} = 1.86T_c = 9.04$ T for Nb$_2$TiW an $\mu_0H^{Pauli} = 1.86T_c = 5.99$ T for Nb$_2$TiMo derived from the Pauli limit effect are both larger than the upper critical field [47].

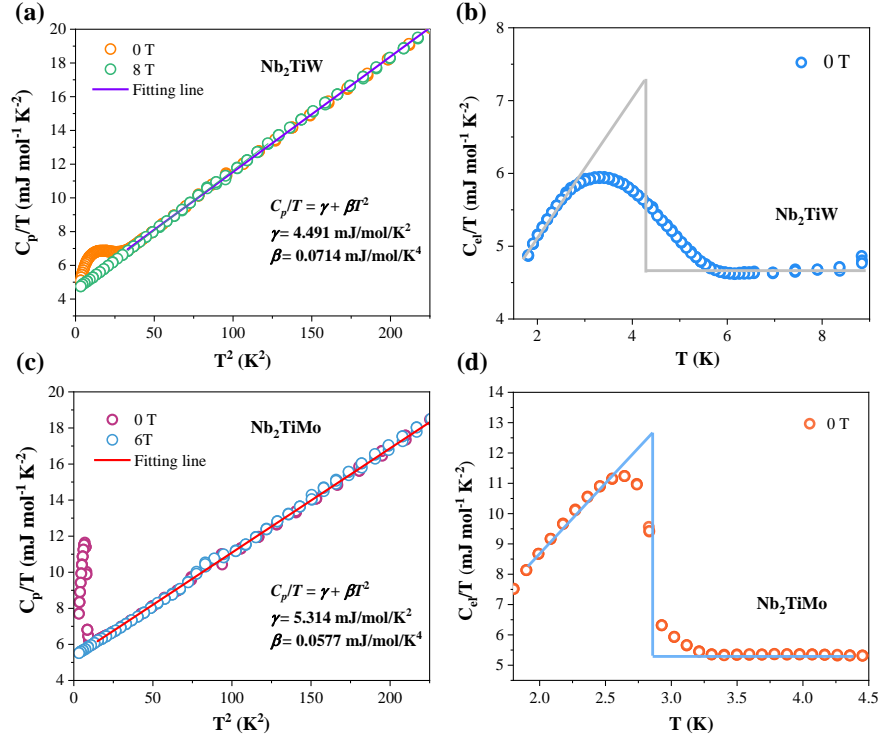

Fig. 6. Specific heat curves of Nb$_2$TiW MEA and Nb$_2$TiMo MEA at zero magnetic field. (a) Nb$_2$TiW MEA specific heat coefficient $C_p/T$ as a function of T$^2$. (b) Nb$_2$TiW MEA electron contribution to heat capacity as a function of $C_{el}/T$. (c) Nb$_2$TiMo MEA specific heat coefficient $C_p/T$ as a function of T$^2$. (d) Nb$_2$TiMo MEA electron contribution to heat capacity as a function of $C_{el}/T$.

The bulk superconducting states of Nb$_2$TiW MEA and Nb$_2$TiMo MEA are further supported by heat capacity measurements, as shown in Fig. 6. In the normal state above $T_c$, the data points can be fitted by the equation $C_p/T = \gamma_n + \beta T^2$, where $\gamma_n$ denotes the Sommerfeld constant for the normal state, and $\beta$ denotes the specific heat coefficient of the lattice part. The fitted data obtained from this equation for Nb$_2$TiW and Nb$_2$TiMo are shown in Fig. 6(a) and Fig. 6(c) for $\gamma_n$(W) = 4.681 mJ/mol/K$^2$ and $\beta$(W) = 0.0685 mJ/mol/K$^4$ and $\gamma_n$(Mo) = 5.314 mJ/mol/K$^2$ and $\beta$(Mo) = 0.0577 mJ/mol/K$^4$, respectively. Fig. 6(b) and Fig. 6(d) show the $C_{el}/T$ - $T$ curves for the temperature ranges of 1.8 K to 10 K and 1.8 K to 4.5 K, respectively, where the $C_{el}$ is obtained from the equation $C_{el} = C_p - \beta T^3$. Estimates determined from the equal-area entropy structure are consistent with $T_c$ extracted from resistivity and magnetization measurements. The Nb$_2$TiMo normalized specific heat jump value $\Delta C_{el}/\gamma_n T$ = 1.35(3), slightly below the Bardeen-Cooper-Schrieffer (BCS) weak coupling limit value of 1.43 for bulk SC. The $Nb_2TiW$ normalized specific heat jump value $\Delta C_{el}/\gamma_n T = 0.81(2)$. The Debye temperature ($\Theta_D$) was then estimated using the Debye model, and the value in the equation $\Theta_D=(12\pi^4 nR/5\beta)^{1/3}$, where $R$ denotes the gas constant and $n$ denotes the number of atoms in the formula cell. The estimated $\Theta_D$ values for $Nb_2TiW$ and $Nb_2TiMo$ are 304 K and 322 K, respectively. Given the $\Theta_D$ and $T_c$, the semi-empirical McMillan formula was used: $\lambda_{ep}=\dfrac{1.04+\mu^* \ln\left(\frac{\Theta_D}{1.45T_c}\right)}{(1-1.62\mu^*)\ln\left(\frac{\Theta_D}{1.45T_c}\right)-1.04}$.[48] The calculation of the electron-phonon coupling constants $\lambda_{ep} = 0.63$ for $Nb_2TiW$ and $\lambda_{ep} = 0.55$ for $Nb_2TiMo$ with $\mu^* = 0.13$ are obtained. In addition, it is also possible to estimate the value of DOS located in the Fermi energy level $N(E_F)$ by using the formula $N(E_F)=\dfrac{3}{\pi^2 k_B^2(1+\lambda_{ep})}\gamma$ with the values of $\gamma$ and $\lambda_{ep}$. The obtained $N(E_F)$ for the $Nb_2TiW$ and $Nb_2TiMo$ samples are 1.22 states $eV^{-1}$ f.u.$^{-1}$ and 1.46 states $eV^{-1}$ f.u.$^{-1}$, respectively.

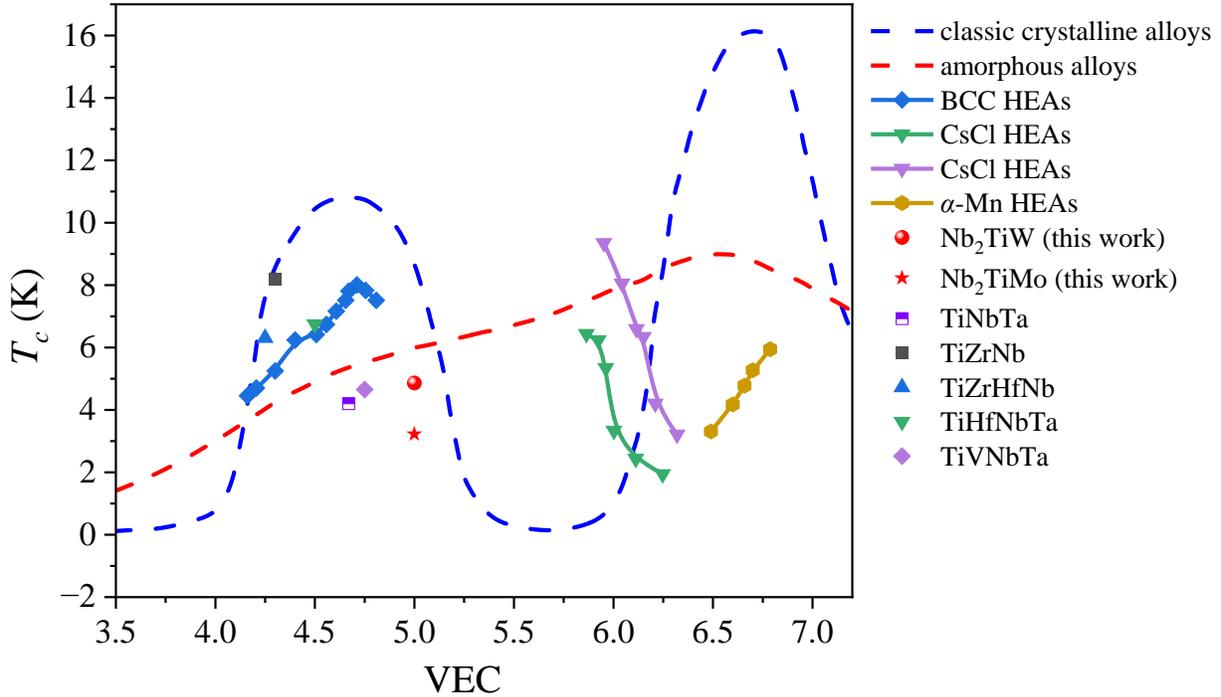

Fig. 7. The phase diagram of VEC-$T_c$ for BCC HEAs, CsCl HEAs, amorphous alloys, classic crystalline alloys, $Nb_2TiW$ and $Nb_2TiMo$ systems.

The phase stability of crystalline solid solutions with slight differences in atomic size is critically influenced by a parameter VEC [5]. For MEA and HEA, calculating the number of valence electrons per alloy is a critical step in designing synthetic superconducting materials. We compare the $T_c$s and VEC of crystalline transition metals and their alloys, amorphous vapor deposition films,

BCC-type HEAs, CsCl-type HEAs and $\alpha$-Mn-type HEAs superconductors to summarize the relationship between them in Fig. 7[17, 18, 20, 23, 27, 28, 49-52]. Nb$_2$TiW and Nb$_2$TiMo are marked with a red dot and a red five-pointed star in the diagram, respectively. The $T_c$ with VEC for Nb$_2$TiW and Nb$_2$TiMo is comparable to other reported bcc-type MEA and HEA superconductors. Notably, the VEC of MEA superconductors plays a crucial role in determining their $T_c$ and material stability. The $T_c$ of MEA superconductors is strongly dependent on the VEC, as seen in the Matthias plots of $T_c$ versus electron counts in simple binary transition metal alloys. Comparing the superconducting transition temperatures of crystalline transition metals and their alloys [53], amorphous deposited thin films [54] and BCC-type HEA superconductors, it is found that the $T_c$s of HEA superconductors is significantly lower than that of the crystalline alloys, and the value shows a monotonically increasing trend higher than that of the amorphous alloys. This suggests that crystallinity has a significant effect on $T_c$ when comparing crystalline binary alloys, HEAs, and amorphous materials at the same electron counts. It is worth noting that a detailed physical comparison between amorphous superconductors, HEA superconductors, and binary superconductors has not yet been performed, but it suggests that the ambiguity of the electron energy ($E$) versus the electron wavevector ($k$) leading to disorder-induced increases in the density of electronic states may be responsible for this behavior. Considering all these results, it can be inferred that a more desirable crystallinity has a positive effect on the $T_c$ of alloys. When considering the $T_c$ of all three superconductors, a trend in all-electron counting can be observed. The small-crystallite BCC-type and CsCl-type HEA superconductors seem to generally follow the bimodal character of the simpler binary alloys with VECs around 4.7 and 5.9, whereas the more complex $\alpha$-Mn-type HEA superconductors do not seem to follow the general trend of the other two systems. Temperature changes induced by atomic substitutions, particularly through the VEC, appear to be a general behavior in HEA superconductors. The $T_c$ of BCC-type superconductors increases with increasing VEC, reaching a maximum value of 7.3 K near 4.7. Similarly, a relationship between the VEC and the $T_c$ has been observed for CsCl-type superconductors, with superconductivity occurring at VECs in the range of 5.9 - 6.3, with a maximum $T_c$ of 9.3 K occurring at a VEC of about 5.9. Therefore, we believe that considering the relationship between VEC and $T_c$ in the exploration of predicting, designing, and synthesizing MEA and HEA superconductors is an effective way to discover new superconducting materials.

To gain more insight into the physical properties of Nb$_2$TiMo and Nb$_2$TiW, we performed DFT calculations to study their electronic structures. Fig. 8 displays the calculated total density of states (DOS) and projected DOS of Nb$_2$TiMo (left panel) and Nb$_2$TiW (right panel). Considering the chemical disorder in the alloys, we employed three different SQSs for each formula, and the average DOS of three SQSs is given in Fig. 8. As shown in the top panel of Fig. 8, the total DOSs of both systems cross the Fermi level, suggesting their metallic properties. The DOSs at the Fermi level are about 2.84 states/eV/f.u. for Nb$_2$TiMo and 2.76 states/eV/f.u. for Nb$_2$TiW. With the calculated DOSs at the Fermi level, the theoretical Sommerfeld constants are estimated to be $\gamma_{Mo} = 6.69\ mJ \cdot mol^{-1} \cdot K^{-2}$ and $\gamma_W = 6.51\ mJ \cdot mol^{-1} \cdot K^{-2}$ through the relationship $\gamma = \frac{1}{3}\pi^2 k_B^2 N_A N(E_F)$, which is close to the experimental values. The local DOS diagrams indicate that Nb and Ti atoms contribute the most to the TDOS near the Fermi level, while Mo or W has a smaller contribution. Furthermore, $d$ orbital plays a dominant role near the Fermi level, namely 3$d$ for Ti, 5$d$ for W, and 4$d$ for Nb and Mo, while $p$ orbital has less contribution, as the projected DOSs reveal. These results imply that the superconductivity may originate from the $d$ electrons of Nb, Ti, and Mo or W, which could introduce strong couplings into the systems.

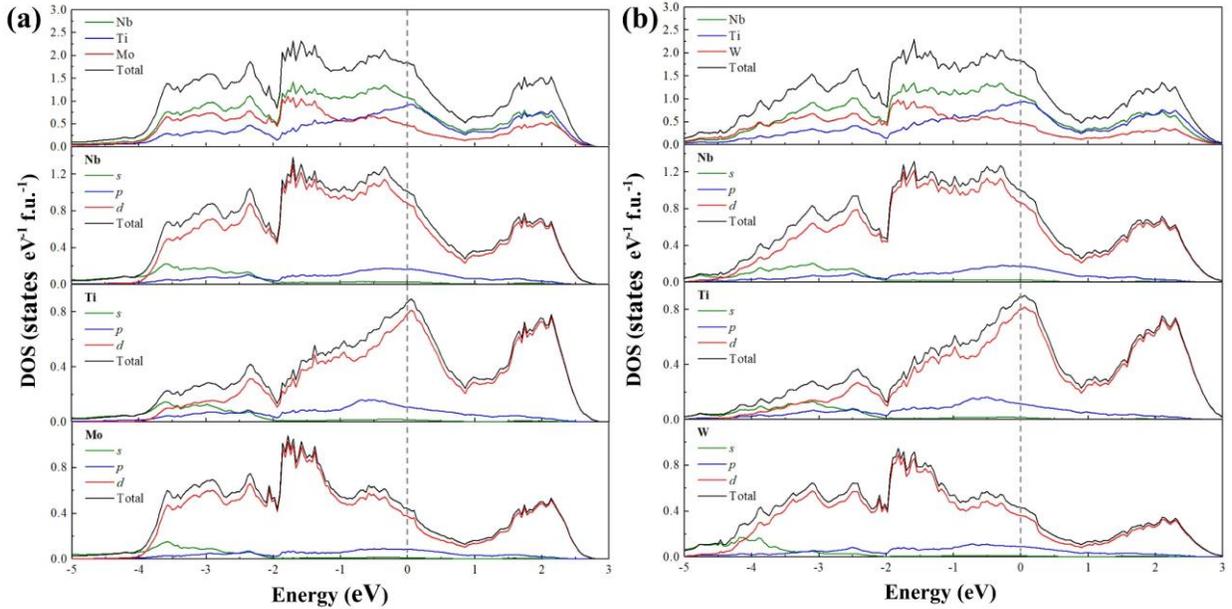

Fig. 8. Local DOS of each element and projected DOS with angular momentum decomposition of each element of Nb$_2$TiMo (a) and Nb$_2$TiW (b). The average DOSs of three structures built by mcsqs (as shown in Fig. S3) are adopted. The gray dashed lines indicate the Fermi level.

In recent years, several HEAs have been reported to host nontrivial band topology [21, 55]. Furthermore, chemical disorder in these HEAs can have influence on the position of Dirac points. Since MEAs share many characterics with HEAs, especially chemical disorder, it is of significance to investigate the topological aspect as well as disorder effect of $Nb_2TiMo$ and $Nb_2TiW$. Therefore, we calculated the band structures of three representative special quasi-random structures (as shown in Fig. S3 in Supporting Information). As Fig. 9 displays, band structures of $Nb_2TiMo$ [Fig. 9 (a-c)] and $Nb_2TiW$ [Fig. 9 (d-f)] show similar features around the Fermi level, indicating that disorder has a modest effect on the electronic properties. Nb and Ti atoms have the largest contributions near the Fermi level, showing consistency with local DOS in Fig. 8. All six structures have dense and complex bands in the vicinity of the Fermi level, resulting in considerable DOSs at the Fermi energy. Note that there are several band-crossing features near the Fermi level, especially around the Z and Γ points, which imply possible topological states. However, as we further analyze, either they split due to the SOC effect originating from $d$ electrons or no band inversion exists around the crossing. Thus, we suppose they are not Dirac points.

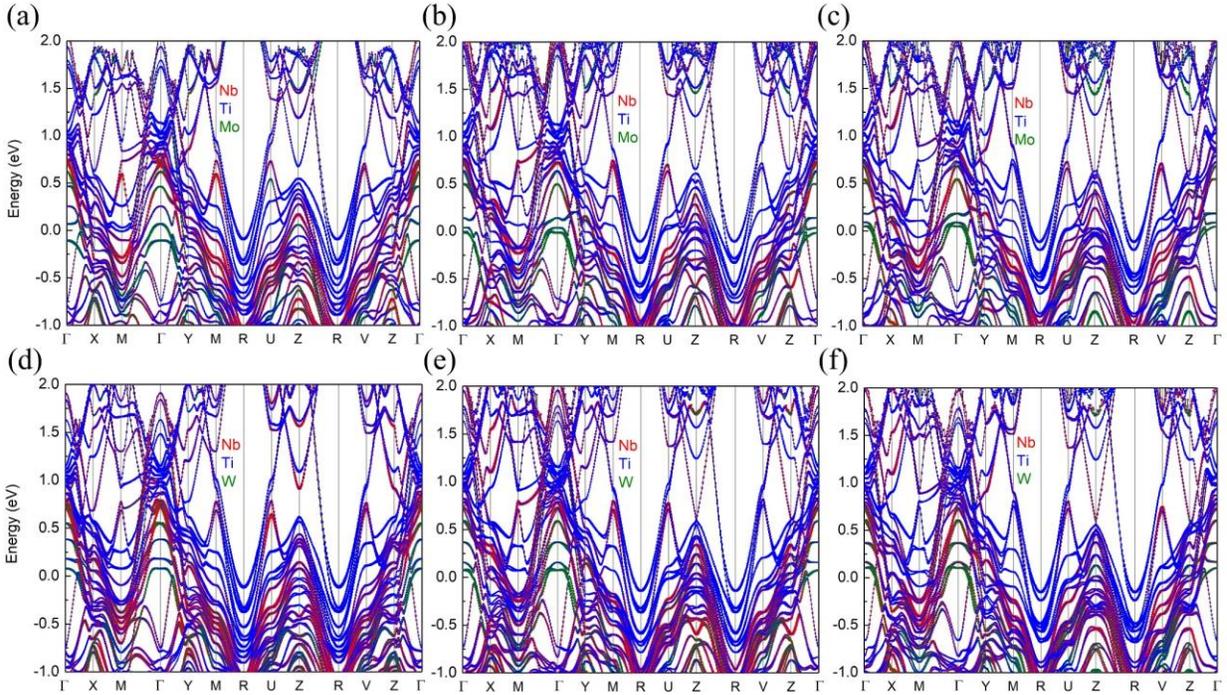

Fig. 9. Band structures of three representative SQSs of $Nb_2TiMo$ (a-c) and $Nb_2TiW$ (e-f) when SOC is considered in DFT calculations. The first, second and third columns are band structures of representative SQSs in Fig. S3 (a)-(c). The red, blue, and olive colors indicate the projection of Nb, Ti, Mo (a-c), and W (e-f). The Fermi level is set to zero.

We have successfully synthesized the new superconductors Nb$_2$TiW and Nb$_2$TiMo MEAs by arc-melting method. XRD results indicated that these two compounds crystallized in BCC structure (Space Group $Im\bar{3}m$, No. 229). The crystal structures of our experimental obtained Nb$_2$TiW and Nb$_2$TiMo compounds differ from the previous theory structures. Resistivity, magnetic and specific heat measurements suggest that Nb$_2$TiW has a $T_c$ of 4.86 K, $\mu_0H_{c1}(0)$ ~ 53.36(2) Oe, and $\mu_0H_{c2}(0)$ ~ 3.52(2) T. Meanwhile, Nb$_2$TiMo has a $T_c$ of 3.22 K, $\mu_0H_{c1}(0)$ ~ 68.23(3) Oe, and $\mu_0H_{c2}(0)$ ~ 2.11(2) T. Nb$_2$TiW and Nb$_2$TiMo MEAs are BCS full-gap $s$-wave superconductors verified by the specific heat measurements. In addition, we studied the band structures of different structural configurations. Similar features were observed near the Fermi level, indicating that the disorder has a moderate effect on the electronic properties. Nb and Ti atoms contribute the most near the Fermi level. By machine learning prediction, Nb$_2$TiW is an ordered structure, crystalizing in $Fm\bar{3}m$ space group. However, both Nb$_2$TiW and Nb$_2$TiMo in our study are MEAs, resulting in the disordered structures. The disorder in MEAs can renormalize electron-phonon/electron-electron interactions, and change the phase coherence of Cooper pairs, and thus suppress superconductivity. Therefore, we suppose the discrepancy of $T_c$ could be attributed to structural disorder. It is noteworthy that these materials are expected to be ductile, making them promising for practical applications in generating strong magnetic fields.

**Knowledgement**

This work is supported by the National Natural Science Foundation of China (12274471, 11922415, 92165204), Guangdong Basic and Applied Basic Research Foundation (2022A1515011168), Guangzhou Basic and Applied Basic Research Project Navigation Project (2024A04J6415). The experiments and calculations reported were conducted at the Guangdong Provincial Key Laboratory of Magnetoelectric Physics and Devices, No. 2022B1212010008. Lingyong Zeng was thankful for the Postdoctoral Fellowship Program of CPSF (GZC20233299) and Fundamental Research Funds for the Central Universities, Sun Yat-sen University (29000-31610058).

**Notes**

The authors declare no competing financial interest.